\documentclass[preprint2]{aastex}

\shorttitle{Detections of C$\_{2}$H, c-C$_{3}$H$_{2}$ and H$^{13}$CN in NGC 1068}
\shortauthors{Nakajima et al.}

\begin{document}

\title{Detections of C$_{2}$H, cyclic-C$_{3}$H$_{2}$, and H$^{13}$CN in NGC 1068}

\author{T. Nakajima\altaffilmark{1} and S. Takano\altaffilmark{1}}
\affil{Nobeyama Radio Observatory, National Astronomical Observatory of Japan, \\
462-2, Nobeyama, Minamimaki, Minamisaku, Nagano, 384-1305, Japan}
\email{nakajima@nro.nao.ac.jp}

\author{K. Kohno\altaffilmark{2,3} and H. Inoue\altaffilmark{2,4}}
\affil{Institute of Astronomy, The University of Tokyo, \\
2-21-1, Osawa, Mitaka, Tokyo, 181-0015, Japan}

\altaffiltext{1}{Nobeyama Radio Observatory, National Astronomical Observatory of Japan}
\altaffiltext{2}{Institute of Astronomy, School of Science, The University of Tokyo}
\altaffiltext{3}{Research Center for the Early Universe, School of Science, The University of Tokyo}
\altaffiltext{4}{Advanced Technology Center, National Astronomical Observatory of Japan}

\begin{abstract}

We used the Nobeyama 45-m telescope to conduct a spectral line survey in the 3-mm band (85.1--98.4 GHz) toward one of the nearest galaxies with active galactic nucleus NGC 1068 and the prototypical starburst galaxy NGC 253. The beam size of this telescope is $\sim$ 18$^{\prime\prime}$, which was sufficient to spatially separate the nuclear molecular emission from the emission of the circumnuclear starburst region in NGC 1068. We detected rotational transitions of C$_{2}$H, cyclic-C$_{3}$H$_{2}$, and H$^{13}$CN in NGC 1068. These are detections of carbon-chain and carbon-ring molecules in NGC 1068. In addition, the C$_{2}$H {\it N} = 1--0 lines were detected in NGC 253. The column densities of C$_{2}$H were determined to be 3.4 $\times$ 10$^{15}$ cm$^{-2}$ in NGC 1068 and 1.8 $\times$ 10$^{15}$ cm$^{-2}$ in NGC 253. The column densities of cyclic-C$_{3}$H$_{2}$ were determined to be 1.7 $\times$ 10$^{13}$ cm$^{-2}$ in NGC 1068 and 4.4 $\times$ 10$^{13}$ cm$^{-2}$ in NGC 253. We calculated the abundances of these molecules relative to CS for both NGC 1068 and NGC 253, and found that there were no significant differences in the abundances between the two galaxies. This result suggests that the basic carbon-containing molecules are either insusceptible to AGN, or are tracing cold ($T_{\rm rot} \sim$ 10 K) molecular gas rather than X-ray irradiated hot gas.

\end{abstract}


\section{Introduction}

To date, about 40 molecular species have been identified in nearby external galaxies (e.g. list available in CDMS\footnote{The Cologne Database for Molecular Spectroscopy (CDMS) can be accessed at http://www.astro.uni-koeln.de/cdms/.}; M\"{u}ller et al. 2005). As a result, it is possible to study molecular abundances and chemical reactions in external galaxies. For example, a difference in molecular abundance has been reported between the nearby starburst galaxies NGC 253 and M 82 (e.g. Mauersberger \& Henkel 1991, Takano et al. 1995), and the reason for this difference has been discussed. They suggested that the temperature in NGC 253 is higher than that in M 82 and/or that the size of the high-temperature region in NGC 253 is much larger than that in M 82. In addition, Takano et al. (2002) pointed out the peculiarity of the molecular abundance in M 82 among nearby starburst galaxies. Some groups have indicated that the specificity of the molecular abundances of M 82 is because of the chemistry of photon-dominated regions (PDRs) (e.g. Garc\'{i}a-Burillo et al. 2002; Mart\'{i}n et al. 2009).

Different galaxies have different properties and activities. These include those with small or large amounts of gas, having a starburst and/or active galactic nucleus (AGN), and with interaction, etc. Molecular line observations of these different galaxies allow us to study the effects of these different properties/activities on the molecular medium. In fact, some groups have suggested that it is possible to diagnose power sources in dusty galaxies using molecular line ratios (e.g. Kohno et al. 2001, Usero et al. 2004, Kohno 2005, Imanishi et al. 2007, Kohno et al. 2008, Krips et al. 2008). The observation of the molecular gas chemistry of the AGN toward NGC 1068, one of the nearest galaxies with an AGN, has been reported by Usero et al. (2004), P\'{e}rez-Beaupuits et al. (2009), and Garc\'{i}a-Burillo et al. (2010). They observed molecular lines of HCO, H$^{13}$CO$^{+}$, SiO, HCO$^{+}$, HOC$^{+}$, and CN, and concluded that the circumnuclear disk of NGC 1068 is a giant X-ray dominated region (XDR). There are theoretical studies on the XDR model by Lepp \& Dalgarno (1996), Maloney et al. (1996), and Meijerink et al. (2005; 2007). However, further systematic observations of molecular lines are indispensable to study the impact of AGN on the interstellar medium. Nevertheless, no systematic unbiased scans of millimeter/submillimeter molecular lines toward AGN have been published to date, although such scans do exist for two nearby starburst galaxies (NGC 253, Mart\'{i}n et al. 2006; M 82, Naylor et al. 2010). Therefore, we started a project to conduct a line survey in the 3-mm band of NGC 1068 using the 45-m telescope at Nobeyama Radio Observatory (NRO) in Japan. The beam size of this telescope (18$^{\prime\prime}$ at 86 GHz) is smaller than the size of the circumnuclear starburst ring in NGC 1068 (d $\sim$ 30$^{\prime\prime}$; e.g. Planesas et al. 1991) , and it is therefore essential to study the impact of the AGN on the surrounding molecules; this will enable us to mitigate the contamination of the molecular lines from the circumnuclear starburst region in NGC 1068. We also observed NGC 253 to compare the effects of AGN on molecular abundance. This is an on-going project, and we present here the initial results of the survey.  

In this study, we report detections of the rotational transition lines of C$_{2}$H, cyclic-C$_{3}$H$_{2}$, and H$^{13}$CN in NGC 1068, and detection of the C$_{2}$H {\it N} = 1--0 transition in NGC 253 as well as other previously reported lines. 

The C$_{2}$H molecule was first detected in interstellar clouds by Tucker et al. (1974). However, extragalactic detections are limited to just a few objects (M82, Henkel et al. 1988; NGC 4945, Henkel et al. 1990; NGC 253, Mart\'{i}n et al. 2006). The C$_{2}$H molecule is important to study the formation and the characteristics of carbon-chain molecules, because it is related to the carbon-chain growth. The cyclic-C$_{3}$H$_{2}$ molecule has been observed in a large number of galactic sources. In external galaxies, it has been detected in Cen A (Seaquist \& Bell 1986; Bell \& Seaquist 1988), NGC 253, and M82 (Mauersberger et al. 1991). Information on the abundances of basic carbon-containing molecules is important to understand the carbon chemistry in NGC 1068. The detection of H$^{13}$CN in the external galaxies has only been reported toward NGC 253 by Mauersberger \& Henkel (1991). It is interesting that the isotope of HCN was detected in NGC 1068, because the isotope ratio gives the constraint on the optical depth of the HCN line; this will help us to unveil the nature of the over-luminous HCN emission at the center of NGC 1068.  

\section{Observations}

The observations at the 3-mm region (85.1 -- 98.4 GHz) toward NGC 1068 and NGC 253 were carried out from February to May, 2009 and from January to May, 2010 using the 45-m telescope. The total observational times under good weather condition are 44.5 and 22.5 hours for NGC 1068 and NGC 253, respectively. The adopted central position and the systemic velocity ($V_{\rm LSR}$) of NGC 1068 are $\alpha_{J2000}$ = 02 h 42 m 40.798 s, $\delta_{J2000}$ = -00$^{\circ}$ 00$^{\prime}$ 47.938$^{\prime\prime}$ and 1150 km s$^{-1}$, respectively (Schinnerer et al. 2000). Similarly, the adopted central position and $V_{\rm LSR}$ of NGC 253 are $\alpha_{J2000}$ = 00 h 47 m 33.3 s, $\delta_{J2000}$ = -25$^{\circ}$ 17$^{\prime}$ 23 $^{\prime\prime}$ (Mart\'{i}n et al. 2006) and 230 km s$^{-1}$ (Takano et al. 1995), respectively. In particular, we focused on the cyclic-C$_{3}$H$_{2}$ {\it J$_{K_{a}, K_{c}}$} = 2$_{1,2}$--1$_{0,1}$ ($\nu_{\rm rest}$ = 85.338906 GHz), H$^{13}$CN {\it J} = 1--0 ($\nu_{\rm rest}$ = 86.340167 GHz, {\it F} = 2--1), C$_{2}$H {\it N} = 1--0 ($\nu_{\rm rest}$ = 87.316925 GHz, {\it J} = 3/2--1/2, {\it F} = 2--1), HCN {\it J} = 1--0 ($\nu_{\rm rest}$ = 88.631847 GHz, {\it F} = 2--1), CH$_{3}$OH {\it J}$_{K}$ = 2$_{K}$--1$_{K}$ ($\nu_{\rm rest}$ = 96.741377 GHz, {\it J}$_{K}$ = 2$_{0}$--1$_{0}$ A$^{+}$), and CS {\it J} = 2--1 ($\nu_{\rm rest}$ = 97.980953 GHz) transitions. These rest frequencies were taken from the database of Lovas (2004).

The dual-polarization sideband-separating receiver (Nakajima et al. 2008) was used with typical system noise temperatures including an atmosphere of about 150--300 K, and the image rejection ratio (IRR) was typically better than 10 dB. It was checked by the IRR measurement system (Nakajima et al. 2010), which employs an artificial signal injection from the top of the receiver optics. The half-power beam widths (HPBWs) at 86 GHz were 18.3$^{\prime\prime}$ and 18.4$^{\prime\prime}$ for H-polarization (H-pol.) and V-polarization (V-pol.), respectively, in 2009, and 19.2$^{\prime\prime}$ and 18.9$^{\prime\prime}$ for H-pol. and V-pol., respectively, in 2010. The measured main beam efficiencies ($\eta_{\rm mb}$) were 0.43 $\pm$ 0.02 and 0.42 $\pm$ 0.02 for H-pol. and V-pol., respectively, in 2009, and 0.49 $\pm$ 0.04 and 0.47 $\pm$ 0.03 for H-pol. and V-pol., respectively, in 2010. The backend used consisted of eight digital spectrometers (Sorai et al. 2000), each having a bandwidth of 512 MHz and a resolution of 605 kHz. We used four spectrometers for each polarization, lined up in the direction of frequency. 

The line intensities were calibrated by the chopper wheel method. The telescope pointing was checked by observing the nearby SiO maser sources, $o$-Cet for NGC 1068, and R-Aqr for NGC 253 every $\sim$1.5 hour. The pointing accuracy was better than about 5$^{\prime\prime}$. The position-switching mode was employed for the observations where the reference position was taken at the position with an azimuthal angle difference of +5$^{\prime}$ for both galaxies. The integration time on each position was 20 sec.

\section{Results}

The main results include detections of C$_{2}$H, cyclic-C$_{3}$H$_{2}$, and H$^{13}$CN toward NGC 1068 and the C$_{2}$H {\it N} = 1--0 transition toward NGC 253. In addition, we observed the HCN {\it J} = 1--0 line and marginally detected CH$_{3}$OH ({\it J}$_{K}$ = 2$_{K}$--1$_{K}$) toward NGC 1068. Very recently, Garc\'{i}a-Burillo et al. (2010) reported the detection of the CH$_{3}$OH group of transitions ({\it J}$_{K}$ = 2$_{K}$--1$_{K}$ and {\it J}$_{K}$ = 3$_{K}$--2$_{K}$). Moreover, detection of CS ({\it J} = 2--1) has already been reported by Tacconi et al. (1997). Nonetheless, we detected this transition in NGC 1068 using a single dish telescope. Figures 1 and 2 show the observed line profiles, and Table 1 summarizes the parameters derived from the Gaussian fits to the observed lines. 

\subsection{C$_{2}$H}

The C$_{2}$H {\it N} = 1--0 line consists of 6 hyperfine components ({\it J} = 3/2--1/2, {\it F} = 1--1, {\it F} = 2--1, {\it F} = 1--0 and {\it J} = 1/2--1/2, {\it F} = 1--1, {\it F} = 0--1, {\it F} = 1--0). We detected two fine structure components because of the line blending. Thus, the fitting has been carried out with a comb of Gaussian profiles at the rest frequencies of the hyperfine components (Gottlieb et al. 1983) using the same widths and with fixed optical depths ($\tau$) ratios relative to the main component based on the theoretical relative intensities (Tucker et al. 1974). The width of the components and the optical depth were taken as free parameters. 

The main-beam brightness temperature ($T_{\rm mb}$) can be calculated from the following equation:
\begin{equation}
T_{\rm mb} = \eta_{\rm bf} \{J(T_{\rm ex}) - J(T_{\rm bg})\} (1 - e^{-\tau}),
\end{equation}
where
\begin{equation}
J(T) = \frac{h \nu}{k} \times \frac{1}{{\rm exp}(\frac{h \nu}{k T}-1)}.
\end{equation}
The beam filling factor, $\eta_{\rm bf}$ = $\theta_{\rm s}^{2}$ / ($\theta_{\rm s}^{2}$ + $\theta_{\rm b}^{2}$), accounts for the dilution effect due to the coupling between the source and the telescope beam in the approximation of a Gaussian source distribution of size $\theta_{\rm s}$ (FWHM) that is observed with a Gaussian beam size $\theta_{\rm b}$ (HPBW). The adopted $\theta_{\rm s}$ of NGC 1068 is 4$^{\prime\prime}$, which is assumed from the distribution of HCN (Helfer \& Britz 1995; Krips et al. 2008). The adopted $\theta_{\rm s}$ of NGC 253 is 20$^{\prime\prime}$, which is assumed from the distribution of CS (Mart\'{i}n et al. 2006). We observed the C$_{2}$H lines only in 2009, and $\theta_{\rm b}$ was measured to be $\sim$ 18.4$^{\prime\prime}$ at 86 GHz in 2009. Thus, for the C$_{2}$H observation, the values of $\eta_{\rm bf}$ for NGC 1068 and NGC 253 with the 45-m telescope are 0.045 and 0.54, respectively. We assumed the excitation temperatures ($T_{\rm ex}$) toward NGC 1068 to be 7.6 K, which is derived from the analysis of the rotation diagram of the CS lines (see subsection 3.3), and that toward NGC 253 to be 5.8 K, which is derived from the analysis of the rotation diagram of the C$_{2}$H lines (see after the next paragraph). The employed cosmic background radiation ($T_{\rm bg}$) is 2.7 K. The fitting results were overlaid on the spectra (Figure 2 (a) and (b)). The Gaussian profiles for C$_{2}$H in NGC 253 do not reproduce the observed spectrum well. The peak position of the main component seems to be shifted possibly because of the self-absorption effect. Because of this shift, the overall fitting results are not as good as those of NGC 1068.

We obtained the optical depth of the C$_{2}$H main component ({\it N} = 1--0, {\it J} = 3/2--1/2, {\it F} = 2--1) of NGC 1068 and NGC 253 to be 0.06 and 0.05, respectively. The total of the optical depth of all the C$_{2}$H components in both galaxies are 0.15 and 0.13, respectively. This indicates that the C$_{2}$H {\it N} = 1--0 lines have small ($\ll$ 1) optical depth in NGC 1068 and NGC 253. If the source size of NGC 1068 is changed to 10$^{\prime\prime}$, to be able to see the effect of the assumed source size, then $T_{\rm ex}$ becomes 8.5 K. In this case, the optical depth is calculated to be 0.01. Therefore, the result of the small optical depth of C$_{2}$H is rather robust considering the uncertainty of the assumed source size. The obtained column densities of C$_{2}$H in NGC 1068 and NGC 253 are 3.4 $\times$ 10$^{15}$ cm$^{-2}$ and 1.8 $\times$ 10$^{15}$ cm$^{-2}$, respectively. For this calculation, we adopted a dipole moment of 0.8 debye (Tucker et al. 1974). On the other hand, the column density of 1.2 $\times$ 10$^{15}$ cm$^{-2}$ was reported in NGC 253 from the C$_{2}$H {\it N} = 2--1 line by Mart\'{i}n et al. (2006). Thus, our result compares well to theirs.

Because the C$_{2}$H {\it N} = 2--1 line was observed by Mart\'{i}n et al. (2006), we made a rotational diagram of C$_{2}$H toward NGC 253 (Figure 3). We adopted the basic equation for the rotation diagram (see for example Turner 1991). The fit to the {\it N} = 1--0 (this work) and 2--1 (Mart\'{i}n et al. 2006) transitions of C$_{2}$H toward NGC 253 appears to indicate a low excitation gas of 5.8 $_{\rm -0.6}^{\rm +0.7}$ K. This value is close to the {\it T}$_{{\rm rot}}$ of NO (6 K) and NS (7 K), which are reported by Mart\'{i}n et al. (2006). These molecular lines trace gas of a low-temperature component. 

\subsection{cyclic-C$_{3}$H$_{2}$}

Spectra of cyclic-C$_{3}$H$_{2}$ are shown in Figure 2 (c) and (d) for NGC 1068 and NGC 253, respectively. The 2$_{1,2}$--1$_{0,1}$ transition was clearly detected in both galaxies. Figure 3 shows the rotational diagram toward NGC 253, where we employed the dipole moment of 3.43 debye reported by Kanata et al. (1987) (Brown et al. (1987) reported it to be 3.32 debye). The statistical weight is 3 for the ortho levels and 1 for the para levels due to two equivalent hydrogen nuclei. In this diagram, we assumed an ortho--to--para ratio of 3. The fit to the 2$_{1,2}$--1$_{0,1}$ (ortho; this work) and 3$_{2,2}$--2$_{1,1}$ (para; Mart\'{i}n et al. 2006) transitions of cyclic-C$_{3}$H$_{2}$ toward NGC 253 appears to indicate a low excitation gas with {\it T}$_{{\rm rot}}$ of 7.6$_{\rm -1.3}^{\rm +1.5}$ K. Mart\'{i}n et al. (2006) detected three transitions of the cyclic-C$_{3}$H$_{2}$ lines and obtained {\it T}$_{{\rm rot}}$ of 9 $\pm$ 8 K from their rotational diagram. However, they reported that the lines are blended with other lines except the 3$_{2,2}$--2$_{1,1}$ transition. This is the reason why we used only their 3$_{2,2}$--2$_{1,1}$ transition in our rotation diagram. As a result, a more probable {\it T}$_{{\rm rot}}$ was obtained. The column density of cyclic-C$_{3}$H$_{2}$ toward NGC 253 is derived using the value of the intersection of the linear regression with the y-axis of Figure 3 and the partition function of cyclic-C$_{3}$H$_{2}$ at 7.6 K. For the calculation of the partition function, the energy levels in Vrtileck et al. (1987) were used. The obtained column density is 4.4 $\times$ 10$^{13}$ cm$^{-2}$. For NGC 1068, the obtained column densities are 1.7 $\times$ 10$^{13}$ cm$^{-2}$ ($\theta_{\rm s}$ = 4$^{\prime\prime}$) and 1.8 $\times$ 10$^{13}$ cm$^{-2}$ ($\theta_{\rm s}$ = 10$^{\prime\prime}$). These column densities in NGC 1068 were derived from the above analysis of the rotation diagram assuming {\it T}$_{{\rm rot}}$ of 7.6 K, which is derived from the rotation diagram of CS (see the next subsection).

\subsection{CS}

Figure 1 (d) shows the CS {\it J} = 2--1 spectrum toward NGC 1068. In addition, the CS {\it J} = 3--2, 5--4, and 7--6 transitions have already been reported by Bayet et al. (2009) in NGC 1068. Using these 4 lines and assuming $\theta_{\rm s}$ of 4$^{\prime\prime}$, the rotational temperature was calculated to be 12.6 K from the rotation diagram as inducted by a dashed line (Figure 3). However, Bayet et al. mentioned that the {\it J} = 7--6 line is marginally detected. Subsequently, with the exception of the {\it J} = 7--6 line, the performed fits are shown by the thin solid line in Figure 3, and the obtained {\it T}$_{\rm rot}$ is 7.6 K. The result of the fitting, excluding the {\it J} = 7--6 line, is deemed appropriate, because the fitting error is significantly small. If the $\theta_{\rm s}$ of NGC 1068 is changed to 10$^{\prime\prime}$, then the {\it T}$_{{\rm rot}}$ will be 13.0 K and 8.5 K (excluding the {\it J} = 7--6 line). We find that the slope of the linear regression in the CS rotational diagram with our new data point is still consistent with previous results (Bayet et al. 2009), indicating that physical properties of the gas traced by CS ({\it J} = 2--1) are similar to those traced by the higher {\it J} transitions of CS. {\it T}$_{{\rm rot}}$ of 7.1 K (excluding the {\it J} = 7--6 line) has been reported by Bayet et al. (2009), and our result is consistent with their result.

\subsection{HCN and H$^{13}$CN}

The relative intensities of the H$^{13}$CN (Figure1 (a)) and HCN (Figure1 (b)) lines can be used as a measure of the optical depth of HCN. The optical depth is derived using the following equation, assuming the same excitation temperature for HCN and H$^{13}$CN, and a $^{12}$C/$^{13}$C isotopic ratio of 50 (Lucas \& Liszt 1998): 
\begin{equation}
\frac{T_{\rm A}^{*}({\rm H^{13}CN})}{T_{\rm A}^{*}({\rm HCN})} = \frac{1-e^{-\tau({\rm HCN})/50}}{1-e^{-\tau({\rm HCN})}}.
\end{equation}
We obtained an optical depth of about 2.6 for the HCN {\it J} = 1--0 line toward NGC 1068. Thus, the HCN {\it J} = 1--0 line in NGC 1068 is optically thick.

\section{Discussion}

It is interesting to examine the effects of AGN on molecular abundance. For example, is there a significant difference in the molecular abundance in circumnuclear disk because of the presence of an AGN? Here, we focused on the C$_{2}$H and cyclic-C$_{3}$H$_{2}$molecules that were detected in this study toward NGC 1068. The relative abundances of C$_{2}$H were calculated with respect to CS in both galaxies to study the effects of the AGN. CS was employed to calculate the relative abundances because the multi-transition data of CS lines have been published (Bayet et al. 2009). In addition, the critical density of CS is about 10$^{4-5}$ cm$^{-3}$, and it traces the region where the densities are higher than that traced by CO. The column densities of CS toward NGC 1068 and NGC 253 were obtained as reported in subsection 3.3.

For C$_{2}$H, we obtained the abundances relative to CS of 5.5 ($\theta_{\rm s}$ = 4$^{\prime\prime}$) and 5.7 ($\theta_{\rm s}$ = 10$^{\prime\prime}$) in NGC 1068, and 4.3 in NGC 253. For cyclic-C$_{3}$H$_{2}$, we obtained the abundances relative to CS of 0.03 ($\theta_{\rm s}$ = 4$^{\prime\prime}$) and 0.16 ($\theta_{\rm s}$ = 10$^{\prime\prime}$) in NGC 1068, and of 0.10 in NGC 253. Mart\'{i}n et al. (2006) reported the column density in NGC 253 to be 3.0 $\times$ 10$^{13}$ cm$^{-2}$, and a relative abundance calculated to be 0.07 in the same way as shown above. It is almost equal in ratio for both of the molecules, and we found no significant differences in the relative abundances of NGC 1068 and NGC 253. Thus, there is a possibility that these basic carbon-containing molecules are insusceptible to AGN. Another possibility is that C$_{2}$H and cyclic-C$_{3}$H$_{2}$ mainly exist in an extended gas away from AGN, because the rotation temperature of C$_{2}$H (7.6 K) is lower than the kinetic temperature of the gas around the central few hundred parsecs ($\geq$ 70 K; Tacconi et al. 1994). However, as far as we know, there is no theoretical study on C$_{2}$H and cyclic-C$_{3}$H$_{2}$ in XDR. Therefore, such model calculations are essential.

\section{Conclusions}

We started line surveys in the 3-mm band toward NGC 1068 and NGC 253. The main results include detections of C$_{2}$H, cyclic-C$_{3}$H$_{2}$, and H$^{13}$CN toward NGC 1068 and the C$_{2}$H {\it N} = 1--0 transition toward NGC 253. We calculated the relative abundances of C$_{2}$H and cyclic-C$_{3}$H$_{2}$ with respect to CS in both galaxies and found no significant differences in the results between NGC 1068 and NGC 253. Thus, it is concluded that these basic carbon-containing molecules are insusceptible to AGN and/or these molecules exist in a cold gas away from AGN. 

We continue an unbiased line survey toward NGC 1068 and NGC 253 and examine the abundances of other molecules in order to study the effects of AGN on the interstellar medium. A clear separation of the circumnuclear disk from the starburst ring is important to understand the abundances of the molecular gas around AGN. With the advent of ALMA, we will be able to study the AGN chemistry with much higher spatial resolution.

\acknowledgments

This work is based on observations with the NRO 45-m telescope, as a part of the line surveys of the legacy projects. The authors thank the project members for helpful discussion. We also thank all of the staff of the 45-m telescope for their support. We are also grateful to Ryohei Kawabe for his advice and support to the line survey project.  

\medskip

Note : After the submission of this manuscript, we found two relevant papers. Snell et al. detected C2H in NGC 1068 (astro-ph 1101.0132).  It was very recently published in The Astronomical Journal 141 38. Costagliola et al. detected C2H and H13CN in NGC 1068 (astro-ph 1101.2122).

\clearpage

\begin{figure}
\plotone{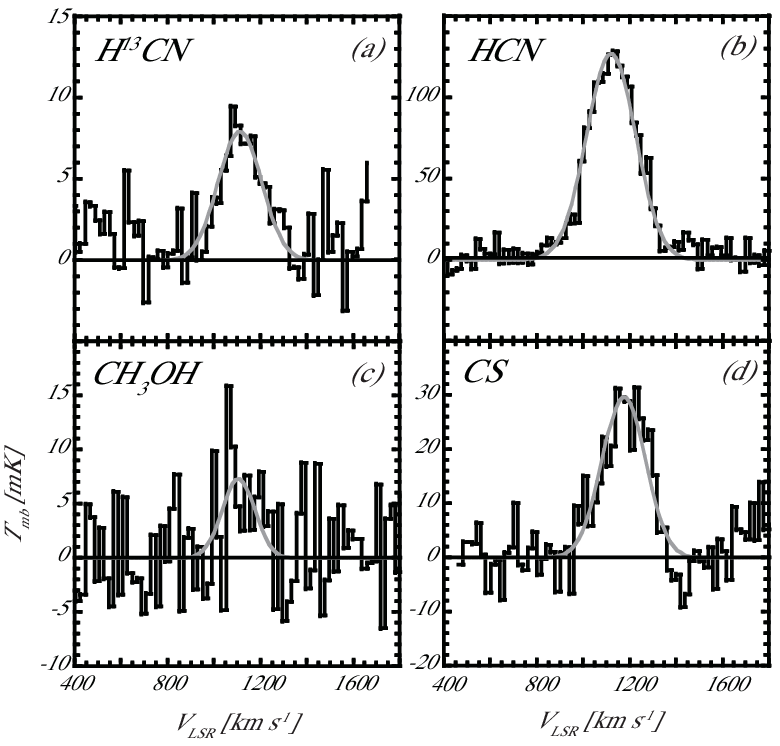}
\caption{(a) H$^{13}$CN {\it J} = 1--0, (b) HCN {\it J} = 1--0, (c) CH$_{3}$OH {\it J}$_{K}$ = 2$_{K}$--1$_{K}$, and (d) CS {\it J} = 2--1 lines toward NGC 1068 obtained with a velocity resolution of 20 km s$^{-1}$. The first order baselines were subtracted. The reference frequencies are listed in the section of observations. The gray curves overlaid on the profiles are the results of Gaussian least-squares fits. Note that the CH$_{3}$OH line is at best tentatively detected (see text). \label{fig1}}
\end{figure}

\begin{figure}
\plotone{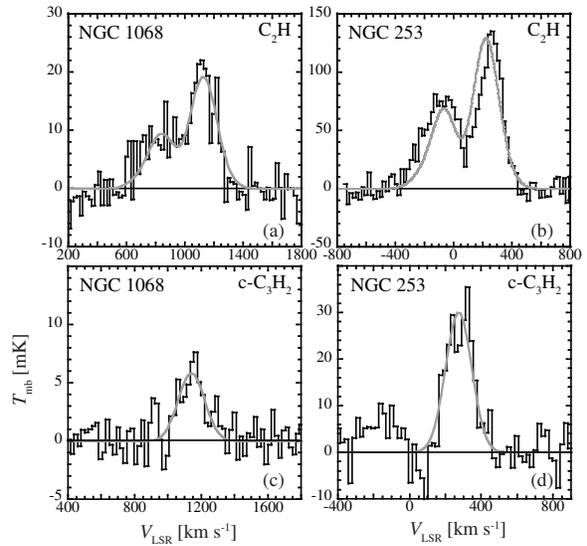}
\caption{The C$_{2}$H {\it N} = 1--0 lines toward (a) NGC 1068 and (b) NGC 253, and the cyclic-C$_{3}$H$_{2}$ {\it J$_{K_{a}, K_{c}}$} = 2$_{1,2}$--1$_{0,1}$ lines toward (c) NGC 1068 and (d) NGC 253. For details, see the caption of Figure 1.\label{fig2}}
\end{figure}

\begin{figure}
\plotone{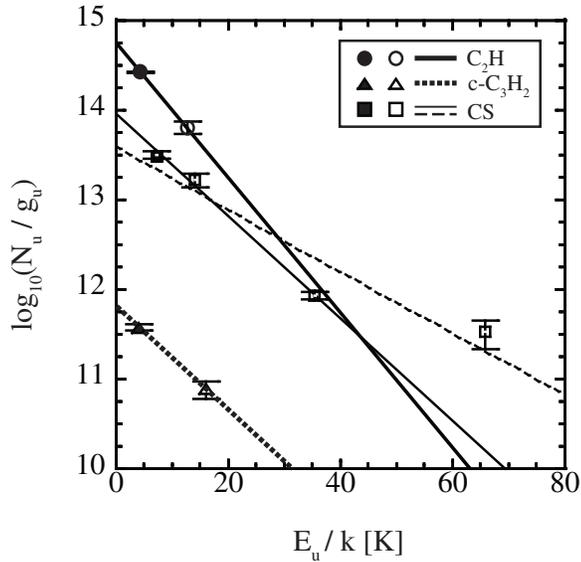}
\caption{Rotational diagrams of C$_{2}$H, cyclic-C$_{3}$H$_{2}$ in NGC 253, and CS in NGC 1068. The black and open circle symbols represent the results of C$_{2}$H of this work and the {\it N} = 2--1 transition by Mart\'{i}n et al. (2006), respectively. The black and open triangle symbols represent the results of cyclic-C$_{3}$H$_{2}$ of this work and the 3$_{2,2}$--2$_{1,1}$ transition by Mart\'{i}n et al. (2006), respectively. The black and open square symbols represent the results of CS of this work and the {\it J} = 3--2, 5--4, and 7--6 transitions by Bayet et al. (2009). The fit with all the data of CS is shown by the dashed line, and the fit excluding the data point of the {\it J} = 7--6 line, is shown by the thin solid line.\label{fig3}}
\end{figure} 

\clearpage

\begin{table}
\begin{center}
\caption{The line parameters\label{tbl-1}}
\begin{tabular}{ccccccc}
\tableline\tableline
line & transition & $\int{T_{\rm mb}{\rm d}V}$ & $T_{\rm mb}$ & $V_{\rm LSR}$ & $\Delta v$(FWHM) & rms  ($T_{\rm mb}$) \\
& & [mK km s$^{-1}$] & [mK] & [km s$^{-1}$] & [km s$^{-1}$] & [mK]\\
\tableline
NGC 1068 &&&&&&\\
cyclic-C$_{3}$H$_{2}$ & {\it J$_{K_{a}, K_{c}}$} = 2$_{1,2}$--1$_{0,1}$ & 1200 & 5 & 1148 & 248 & 1.7\\
H$^{13}$CN & {\it J} = 1--0 & 1900 & 9 & 1111 & 234 & 4.1\\
C$_{2}$H & {\it N} = 1--0 & 6900 & 19 & 1139 & 209 & 3.3\\
HCN & {\it J} = 1--0 & 35000 & 128 & 1117 & 336 & 19.1\\
CH$_{3}$OH & {\it J}$_{K}$ = 2$_{K}$--1$_{K}$ & 1500 & 7 & 1104 & 235 & 3.8\\
CS & {\it J} = 2--1 & 6400 & 30 & 1174 & 312 & 4.8\\
\tableline
NGC 253 &&&&&&\\
cyclic-C$_{3}$H$_{2}$ & {\it J$_{K_{a}, K_{c}}$} = 2$_{1,2}$--1$_{0,1}$ & 5500 & 30 & 272 & 243 & 4.6\\
C$_{2}$H & {\it N} = 1--0 & 48000 & 129 & 219 & 199 & 9.1\\
\tableline
\end{tabular}
\end{center}
\end{table}

\end{document}